\documentclass[10pt]{iopart}

\newcommand{\figtwob}{2.1in}
\newcommand{\figtwo}{2.3in}

\usepackage{iopams}  
\usepackage{graphics}
\usepackage{setstack}
\usepackage[abbr]{harvard}
\usepackage{epsf}
\usepackage{psfrag}
\usepackage[pagewise,running]{lineno}
\usepackage{graphicx}
\usepackage[squaren]{SIunits}
\usepackage{psfig}

\def\eps{\varepsilon}
\def\epsinfa{\epsilon}

\begin{document}
 \submitted{\JPD}

%\begin{pagewiselinenumbers}

  \article[Disordered binary dielectric mixtures (50-50)]{
%    \begin{flushright}
%      23-27 March Heriot-Watt University\\ {\em The} \textsf{Physics} 
%      Congress 2003\\ Dielectrics For Emerging Technologies    
%    \end{flushright}
  }
  {Signs of low frequency dispersions in disordered 
    binary dielectric mixtures (50-50)\footnote[2]{Financed by ELIS project of the Swedish Foundation for Strategic Research (SSF)}}

  \author{Enis Tuncer\footnote[1]{Present address: Applied Condensed-Matter Physics, 
      Department of Physics, University of Potsdam, Am Neuen Palais 10,
      D-14469 Potsdam, Germany (enis.tuncer@physics.org)}
  }

\address{High Voltage Division, Department of Electric Power Engineering, Chalmers University of Technology, SE-412\ 96 G{\"o}teborg, Sweden}

\begin{abstract}Dielectric relaxation in disordered dielectric mixtures are presented by emphasizing the interfacial  polarization. The obtained results coincide  with and cause confusion with those of the low frequency dispersion behavior. The considered systems are composed of two phases on two-dimensional square and triangular topological networks. We use the finite element method to calculate the effective dielectric permittivities of randomly generated structures. The dielectric relaxation phenomena together with the dielectric permittivity values at constant frequencies are investigated, and significant differences of the square and triangular topologies are observed. The frequency dependent properties of some of the generated structures are examined. We conclude that the topological disorder may lead to the normal or anomalous low frequency dispersion if the electrical properties of the phases are chosen properly, such that for ``slightly'' {\em reciprocal mixture}--when $\sigma_1\gg\sigma_2$, and $\epsilon_1<\epsilon_2$--normal, and while for ``extreme'' {\em reciprocal mixture}--when $\sigma_1\gg\sigma_2$, and $\epsilon_1\ll\epsilon_2$--anomalous low frequency dispersions are obtained. Finally, comparison with experimental data indicates that one can obtain valuable information from simulations when the material properties of the constituents are not available and of importance.
%the former dispersion is obtained for $\epsilon_1<\epsilon_2$. Similarly, for $\epsilon_1\ll\epsilon_2$ and $\sigma_1\gg\sigma_2$ the dielectric response of the systems show the  anomalous low frequency dispersion behavior.\\
%{\bf Key Words} {Disordered solids}, {Dielectric properties of solids and liquids}, {Composite materials}, {Complex dielectric constant}, Disordered solids electrical conductivity
\end{abstract}
%\maketitle

\section{Introduction}
Numerical simulations and analytical formulas on dielectric behavior of materials with two different phases (binary mixtures) on regular two dimensional structures have been shown to exhibit dielectric behavior of Debye type when the concentration, $q$, of inclusion phase is low~\cite{Tuncer2001a,TuncerReview}. The polarization relaxation appears due to the differences between the electrical properties of the constituents, $\epsilon_1/\sigma_1\ne\epsilon_2/\sigma_2$. This polarization is called Maxwell-Wagner-Sillars ({\sc mws})~\cite{Maxwellbook,Wagner1914,Sillars1937} polarization (also known as the interfacial polarization).  At higher concentrations of inclusions the dielectric behavior deviates slightly from Debye type~\cite{Tuncer2001a} when the mixture is a match-composite--the electrical properties of phases, permittivity $\varepsilon$ and direct-current  (dc) conductivity $\sigma$, are chosen in such a way that $\varepsilon_1<\varepsilon_2$ and  $\sigma_1<\sigma_2$ where the subscripts (1) and (2) indicate the matrix and inclusion phases. The deviation from  Debye type increases for reciprocal-composites--when $\varepsilon_1<\varepsilon_2$ and  $\sigma_1>\sigma_2$~\cite{Tuncer2002b}. Previously, {\sc mws} polarization in disordered structures have been reported~\cite{Tuncer2002b}, and non-Debye dielectric relaxations have been observed. 

In this paper, we stress on the dielectric properties of disordered binary mixtures, and extend the investigations on mixtures to micro-structural differences  and importance of ratio of electrical properties of constituents. The electrical properties of the structures were calculated by using the finite element method ({\sc fem})~\cite{TuncerAcc1}. The dielectric responses of several selected disordered systems are presented. The dielectric permittivity values calculated were as high as in the porous solids~\cite{Bo1996,Papathan2000,Papathan2001,Capaccioli3,shen1990}. Therefore, the results of two dimensional calculations are compared not only to dielectric responses of binary polymeric composites~\cite{Tuncer2000a,Tuncer2001e} and of porous liquid-solid systems as well.

\section{Low frequency dispersion}
Dielectric properties of materials can be characterized by the complex dielectric permittivity, 
$$\varepsilon(\omega)=\varepsilon'(\omega)-\imath\varepsilon''(\omega)$$ 
or in susceptibility notation, 
$$\chi(\omega)\equiv\varepsilon(\omega)-\epsilon$$ 
which have the angular frequency $\omega$ dependencies. Here, the instantaneous dielectric permittivity $\epsilon$ consists of the relaxations which are faster than the relaxations observed in the considered frequency range. External variables, such as, temperature, pressure, humidity {\em etc.,} also influence the dielectric properties of materials. When a relaxation process is present, the dielectric response of materials usually shows a peak in the frequency dependence of the imaginary part of the loss component, $\chi''(\omega)$, and a dielectric increment in the real component, $\chi'(\omega)$~\cite{Jonscher1983}. The response is often of non-Debye type~\cite{NonDebye}, and can in general be expressed by  Havriliak-Negami empirical formula~\cite{HN},  
\begin{equation}
  \label{eq:HN}
  \varepsilon(\omega;\epsilon,\chi(0),\tau,\alpha,\beta)=\epsilon+\frac{\chi(0)}{[1+(\imath\omega\tau)^\alpha]^\beta}
\end{equation}
where %$\epsinfa$ is the instantaneous dielectric polarization, and $\imath=\sqrt{-1}$. 
$\tau$ and $\chi(0)$ represent the average relaxation time (inverse relaxation rate) and the dielectric strength of the polarization, respectively. $\alpha$ and $\beta$ are parameters that display the degree of deviation from the Debye relaxation, $\alpha\le1\ \land\ \alpha\beta\le1$. When $\alpha=\beta=1$, Eq.~(\ref{eq:HN}) yields Debye's relaxation equation~\cite{Debye1945}. The other empirical expressions that can be obtained from Eq.~(\ref{eq:HN}) are when $\beta=1$, Cole-Cole~\cite{CC} and when $\alpha=1$ Davidson-Cole~\cite{CD}, respectively. 
\begin{figure}[t]
  \centering{\includegraphics[height=3.5in,angle=-90]{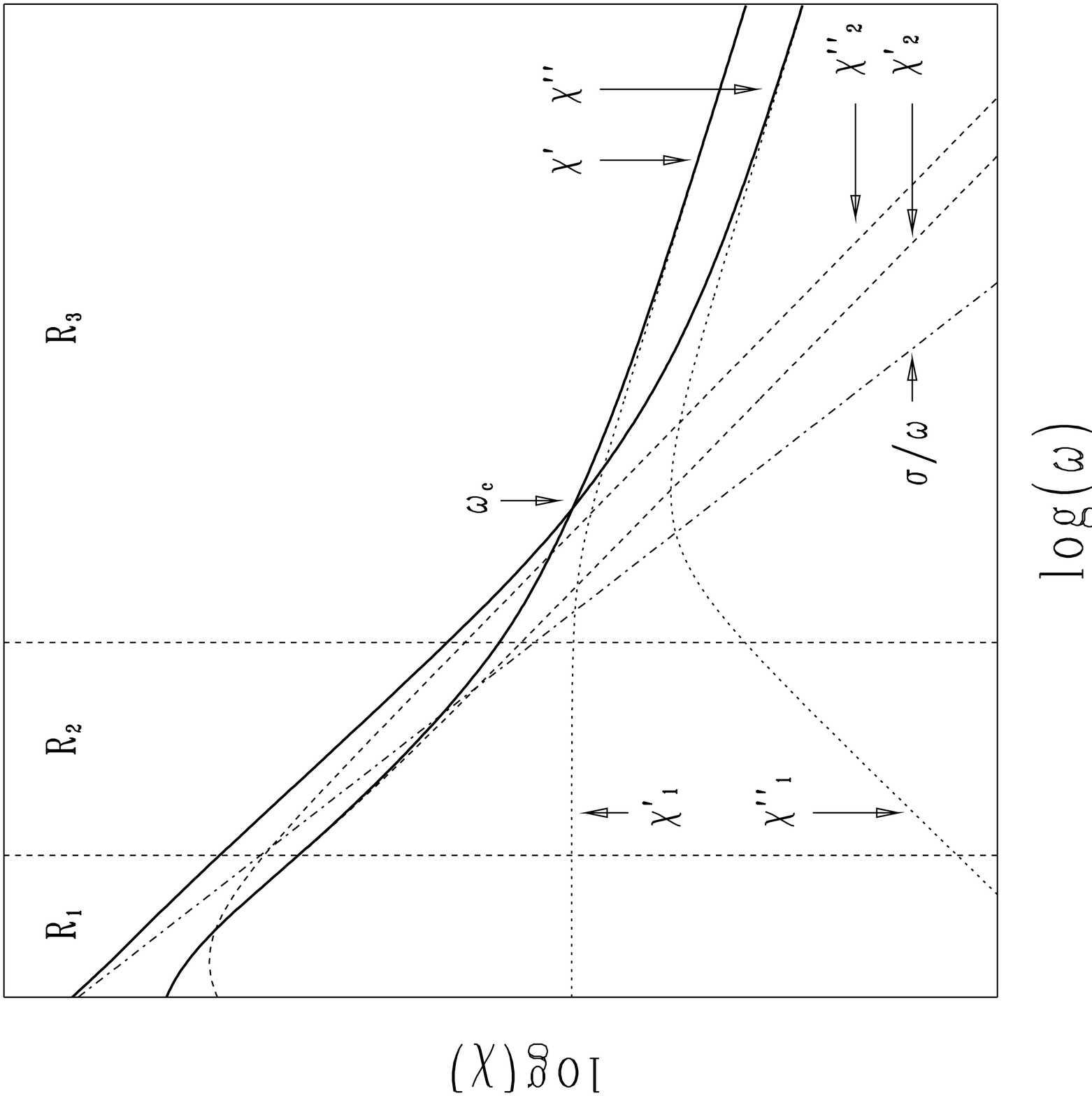}}
  \caption{Anomalous low frequency dispersion which is composed of two responses, $\chi_{1,2}$, as in Eq.~(\ref{eq:HN}) and a dc contribution, $\sigma/\omega$. The thick solid line ($\full$) is the total response. The dashed ($\dashed$) and broken ($\broken$) lines are the relaxation polarizations. The chain line ($\chain$) is the dc conduction contribution.  \label{fig:lfd}}
\end{figure}

At frequencies much lower and much higher than the inverse relaxation time, $\tau^{-1}$, the susceptibility can be written as
\begin{equation}
  \label{eq:non-Debye1}
  \chi'(\omega)\propto\chi''(\omega)\propto\omega^{-\alpha\beta}\quad \omega\gg\tau^{-1}
\end{equation}
\begin{equation}
  \label{eq:non-Debye2}
  \chi''(\omega)\propto1-\chi'(\omega)/\chi(0)\propto\omega^{\alpha}\quad \omega\ll\tau^{-1}
\end{equation}
Although Eqs.~(\ref{eq:HN}) and (\ref{eq:non-Debye1}) are applicable for dielectric responses at high frequencies, at low frequencies the peak of the response is not often observable; the relaxation process is not finalized yet and measured data disregard Eq.~(\ref{eq:non-Debye2}) for those processes. Eq.~(\ref{eq:non-Debye1}) describes the ``normal'' low frequency dispersion or an empirical dielectric function, also known as constant-phase element~\cite{MacDonald1987}.
\begin{equation}
  \label{lfd}
  \chi=A(\imath\omega)^{a-1}\quad\text{and}\quad a<1 
\end{equation}
At the lowest possible frequencies, it is often observed that there is a transition of the exponent, $\alpha\beta$ or $a$, from a low value at high frequencies to a higher value at low frequencies beyond a characteristic frequency $\omega_c$ is passed~\cite{Jonscher1983,HillLFD}. This phenomenon is called the `anomalous' low frequency dispersion. %Examples of the anomalous low frequency dispersion have been reported in the literature~\cite{HillLFD}. 

One can construct an anomalous low frequency dispersion as the summation of two dielectric relaxations with $\tau_1<\tau_2$ as in the form of Eq.~(\ref{eq:HN}), and furthermore, a dc conduction contribution $\sigma/\omega$ can as well be included, where $\sigma$ is normalized conductivity--divided by the permittivity of the free space, $\epsilon_0\equiv 1/36\pi\ \nano\farad\per\meter$.
\begin{equation}
  \label{eq:twohn}
   \chi_{\sf a}(\omega)=\frac{\chi_1(0)}{[1+(\imath\omega\tau_1)^{\alpha_1}]^{\beta_1}} + \frac{\chi_2(0)}{[1+(\imath\omega\tau_2)^{\alpha_2}]^{\beta_2}}+\frac{\sigma}{\imath\omega}
\end{equation}
 The resulting dielectric response is illustrated in Figure~\ref{fig:lfd}. Now we discuss the different marked-regions in the figure. The real and imaginary parts of the total susceptibilities $\chi_{\sf a}'(\omega)$ and $\chi_{\sf a}''(\omega)$ have the same slopes, but different amplitudes on the frequency regions above, $\omega\gg\omega_c$, and below, $\omega\gg\omega_c$, the critical frequency $\omega_c$. In the figure, we, therefore, divide the frequency region into three, $R_{1},~R_{2}~ \text{and}~ R_{3}$, which may represent a typical experimental data-set. When $R_3$ is considered, the real part of the response, $\chi_{\sf a}'(\omega)\approx \chi_1'(\omega)$, is higher than the imaginary part, $\chi_{\sf a}''(\omega)\approx \chi_2'(\omega)$ which indicates that 
\begin{equation}
  \label{eq:cond1}
\sqrt{(\alpha_1-1)^2+(\beta_1-1)^2}>0.5  
\end{equation}
This response in $R_3$ can be confused with a response containing relaxation polarization and dc conduction contributions as expressed below since the finalization of the polarization is not clear within this finite frequency range.
\begin{equation}
  \label{eq:hn+dc}
  \chi(\omega)=\frac{\chi(0)}{[1+(\imath\omega\tau)^\alpha]^\beta}+\frac{\sigma}{\imath\omega}
\end{equation}
In region $R_2$, $\chi_{\sf a}''(\omega)$ is higher than $\chi_{\sf a}'(\omega)$ which shows that in Eq.~(\ref{eq:twohn})
\begin{equation}
  \label{eq:cond2}
\sqrt{(\alpha_2-1)^2+(\beta_2-1)^2}<0.5  
\end{equation}
The regions $R_3$ and $R_2$ can separately be described by low frequency dispersions~\cite{Jonscher1983}. Consequently, the region $R_3 \cup R_2$ is the anomalous low frequency dispersion~\cite{HillLFD} which is then expressed by summations two power laws;
\begin{equation}
  \label{eq:twopowerlaws}
  \chi(\omega)=A(\imath\omega)^{a-1} + B(\imath\omega)^{-b} 
\end{equation}
where $a$ and $b$ are related to $\alpha$ and $\beta$ as in Eq.~(\ref{eq:HN}), and they satisfy the conditions in Eqs.~(\ref{eq:cond1}) and (\ref{eq:cond2}) in regions $R_2$ and $R_3$. $A$ and $B$ depend on the concentration of charge carriers, and they are thermally activated constants~\cite{MacDonald1987}. The intersection of $\chi_{\sf a}'(\omega)$ and $\chi_{\sf a}''(\omega)$ is the critical frequency, $\omega_c$, and it depends on the response parameters.  

In region $R_1$, the presence of the dc conductivity could be observed significantly, but, still $\chi_{\sf a}''(\omega)\ne\sigma\omega^{-1}$. The dc conductivity contribution hinders the loss peak of $\chi_{2}''(\omega)$.  The bending of or convergence of the real part of the dielectric susceptibility, $\chi_{\sf a}'(\omega)$, indicates that the relaxation polarization is to finalize soon.

In the literature, the low frequency dispersion is associated with materials showing defect dominated conduction (hopping conduction) mechanism. Therefore, the concentration of defects is critical in the sense that local field changes in the materials influences the electrical properties, such as, conductivity of materials that do not obey  Ohm's law~\cite{sor84} in the presence of the interaction of charge carriers with themselves and with the defects. The micro-structure of the considered medium affects the conductivity of the system depending on a length scale which is defined as the distance between the charge scatterer or defects~\cite{HillLFD,sor84}. Moreover, similar length scales can also be defined as the variation of the dielectric permittivity in a continuous random dielectric~\cite{men84} and the variation of charge density or the electric field. All these considerations are valid when there exists some kind of disorder (randomness) in the material, {\em i.e.}, distribution of defects, activation energies, {\em etc.} 
\begin{figure}[t]
  \begin{center}
    \begin{tabular}{cc}
      \psfragscanon
      \psfrag{FRAME16}{}
      {\resizebox{1.34in}{!}{\includegraphics[]{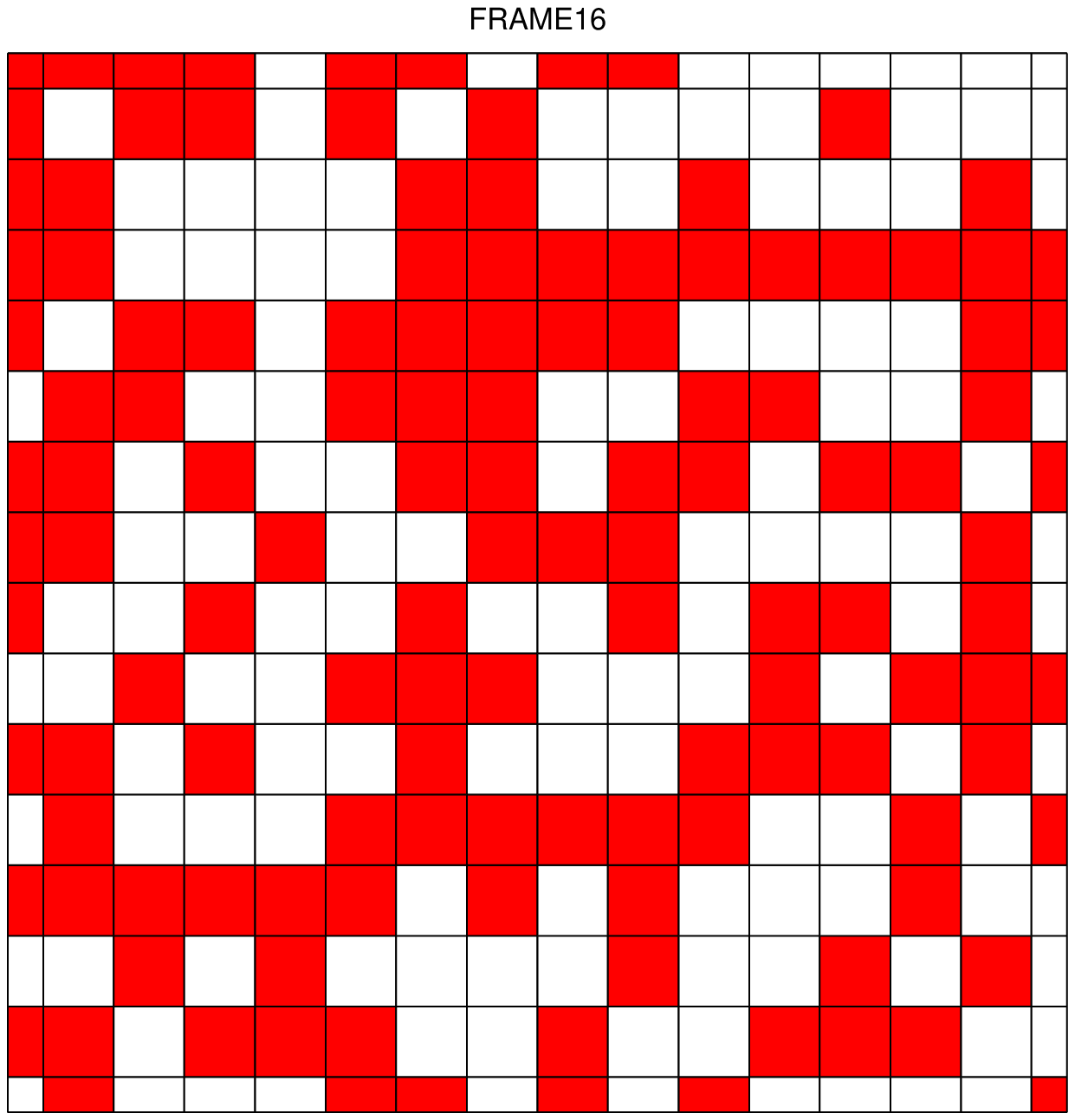}}}
      \psfragscanoff
      &
      \psfragscanon
      \psfrag{FRAME16}{}
      {\resizebox{1.6in}{1.4in}{\includegraphics[]{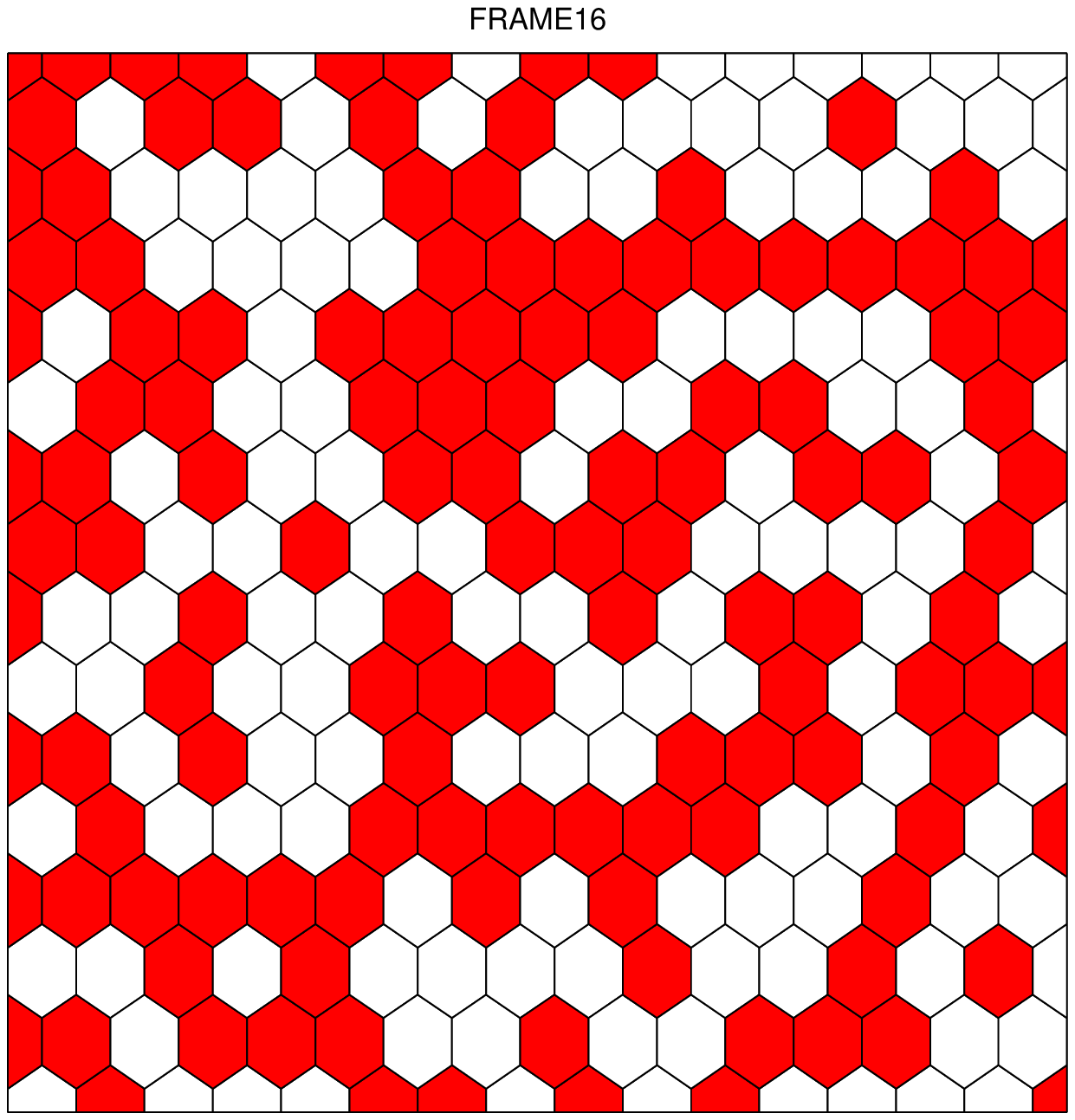}}}
      \psfragscanoff
\\
      (a) &(b) \\ 
    \end{tabular}
  \end{center}
  \caption{\label{fig:geom2}(a) Square and  (b) honeycomb $16 \times 16$ random structures corresponding to the same set of lattice computer topology, ($i,j$). The dark and white polygons are phases with different electrical properties. The outer lines ($\full$) are chosen such that the structures are the computation domains.}
\end{figure}

\section{Modelling considerations}
The electrical properties of the phases, {\em i.e.}, dielectric permittivity and ohmic conductivity are frequency independent, $\eps_{1,2}(\omega)=\epsinfa_{1,2}$ and $\sigma_{1,2}=\text{constant}$, in the calculations. The phases are linear materials, and the dielectric permittivities, $\epsilon_{1}$ and $\epsilon_2$, of the first (1) and the second (1) phases are taken as $2$ and $10$, respectively. The ohmic conductivities of the media are $100\pico\siemens\per\meter \simeq 11.3\ \varepsilon_0$ and $1\pico\siemens\per\meter \simeq 1.13\ \varepsilon_0$. The considered values of the conductivity and dielectric permittivity yield a polarization relaxation around $1\ \hertz$.
The concentration of the phases are fifty-fifty, $q_{1,2}=0.5$.

%Since the dielectric relaxation of regular composite systems are sensitive to the 
Two-dimensional networks of square and triangular topological arrangements are considered. The networks are composed of $16\times16$ cells, as presented in Figure~\ref{fig:geom2}, and for $q_{1,2}=0.5$, the number of possible different topology configurations is $\sim 5.7687\times 10^{75}$. 

Disordered structures are generated on these lattices by randomly assigned material properties such that the distribution of phases are the same for both lattice arrangements. This is achieved by generating distributions in the $i$-$j$ plane, where $i$ and $j$ are integers between $0$ and $15$. The mapping ($x$-$y$ positions) of the cell centers in the square network are, then, expressed as
\begin{equation}
    x=\xi \left ( \frac{1}{2}+i \right ) \qquad y=\xi \left (\frac{1}{2}+j \right) \label{eq:square_mapping}
\end{equation}
where $\xi$ is the length of lattice spacing. In brief, Eq.~(\ref{eq:square_mapping}) defines the physical coordinates $(x,y)$ using the square lattice computer topology, $(i,j)$. In a similar way, since the computer topology of the random structures are assumed to be the same, as illustrated in Figure~\ref{fig:geom2}, the values of $(i,j)$ is used in the triangular lattice topology to assign the physical coordinates $(x,y)$~\cite{Cellular},
\begin{eqnarray}
    \text{Even }j: \qquad x=\xi \left(\frac{1}{2}+i\right ) \qquad y=\xi \frac{\sqrt{3}}{2}\left (\frac{1}{2}+j\right ) \\
    \text{Odd }j: \qquad x=\xi (1+i) \qquad y=\xi \frac{\sqrt{3}}{2}\left(\frac{1}{2}+j\right ) \\
\label{eq:triangle_mapping}
\end{eqnarray}
The inclusions are distributed using a uniform random generator. In this approach, each square in the square-network and each honeycomb in the triangular-network are treated as the smallest parts of the constituents. Since the corners of squares and honeycombs introduce problems in the {\sc fem} calculations, they are rounded without affecting the average shape and area of the inclusions. The rounding process also improves the mesh of the computational domain in the finite element method. We use only 1024 random structures which is only $\sim1.7751\times10^{-73}$ part of the possible topology configurations. 

We do not consider the finite-size effects due to the meshing complexity in the {\sc fem} procedure. The first feature of applying two different networks is related to the site-bond percolation in the considered networks. When $q_{1,2}=0.5$, the area fraction of square network is under the site-percolation threshold ($\sim0.592$) and equal to the bond percolation ($0.5$). At the same concentration, the area fraction, on the other hand, is equal to the site-percolation threshold ($0.5$) and over the bond percolation ($\sim0.347$)~\cite{Percolation}. The second feature of our geometrical modelling includes the electromagnetic field theory which is not taken into consideration in the percolation theory, such as, the conductivity calculations of random structures and networks~\cite{Percolation}.

\section{Results and Discussion}
\subsection{Dielectric permittivities at constant frequencies}
\begin{figure}[t]
  \centering{\includegraphics[height=\figtwo]{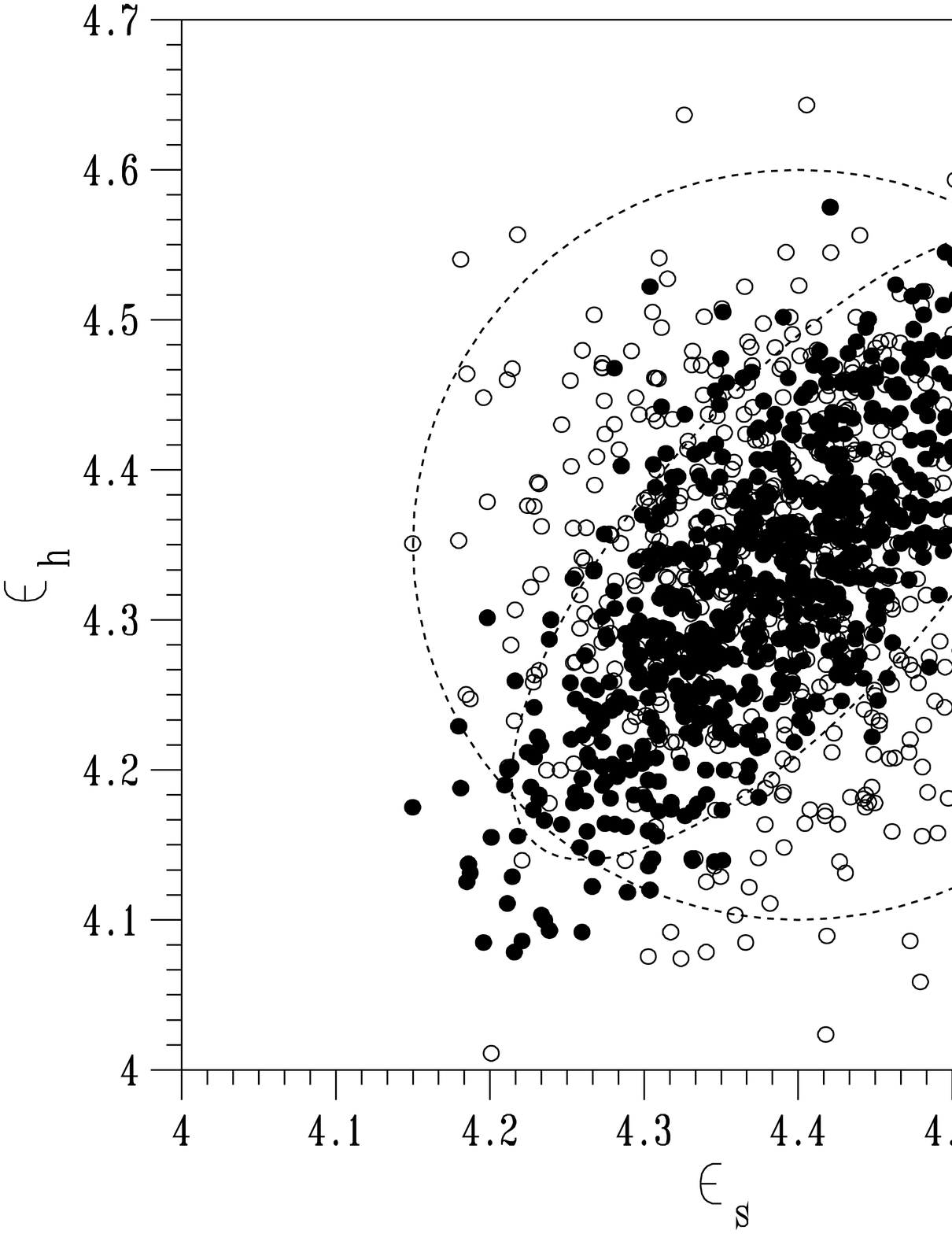}
  \includegraphics[height=\figtwo]{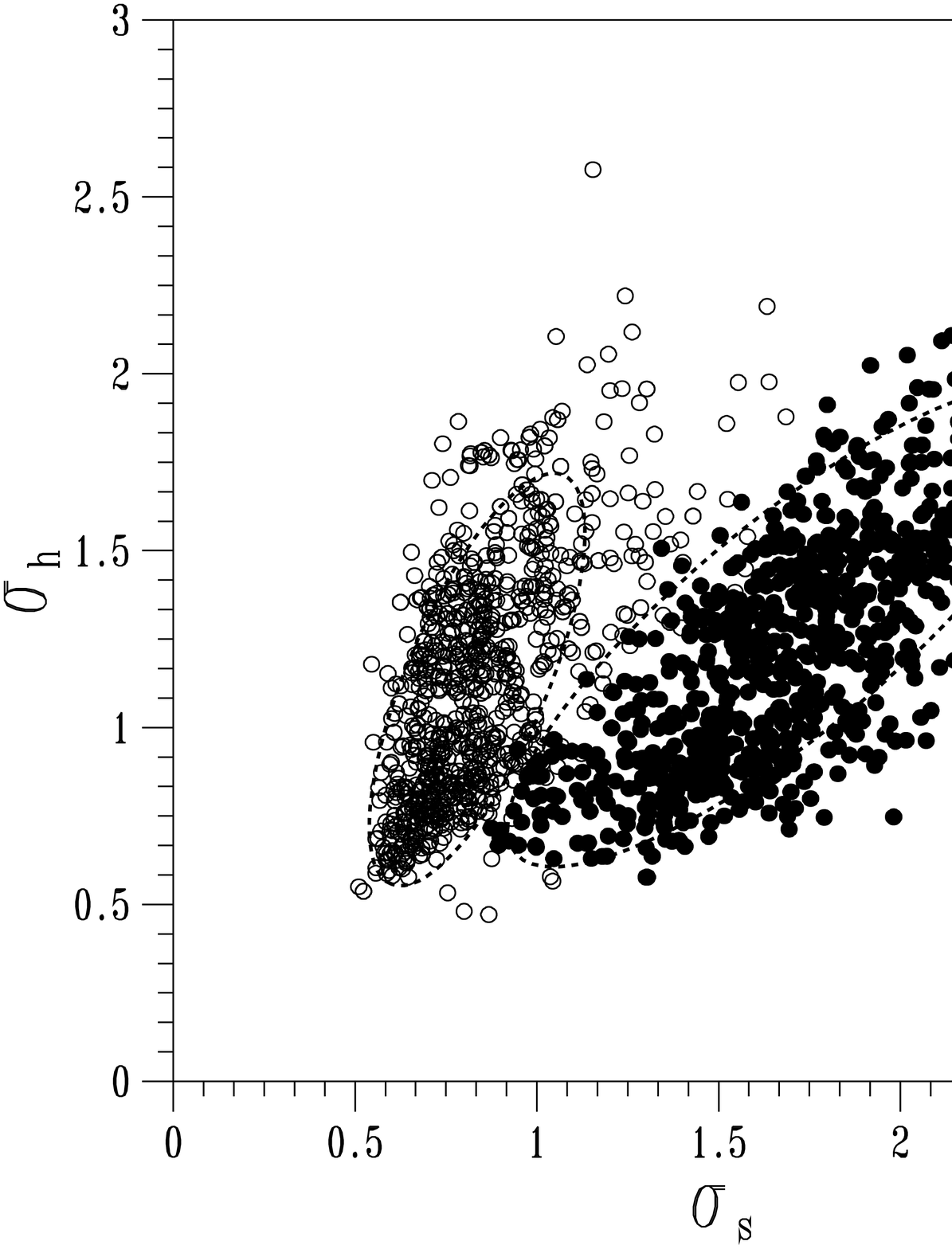}}
  \caption{Relation between (a) permittivities at high frequencies ($\omega=200\pi\ \hertz$), $\epsilon$, and (b) normalized conductivities at low frequencies ($\omega=2\pi\ \milli\hertz$), $\sigma$ for the disordered structures with different networks. Subscripts $h$ and $s$ represent the honeycomb and square tiles. The open ($\circ$) and filled ($\bullet$) symbols represent the considered cases when $\epsilon_1<\epsilon_2$ and $\sigma_1<\sigma_2$ and when $\epsilon_1<\epsilon_2$ and $\sigma_1>\sigma_2$, respectively.  \label{fig:hig}}
\end{figure}
First, the influence of topological differences are taken into consideration by investigating the dielectric permittivities at high ($\omega=200\pi\ \hertz$) and at low ($\omega=2\pi\ \milli\hertz$) frequencies. Both analytical mixture expressions and calculations by assuming non-percolating regular lattice structures yield steady permittivity values, $\epsilon$ and $\epsilon+\chi(0)$, at these frequencies~\cite{Tuncer2001a}. Two different cases are considered in which the permittivities of the phases are kept constant, $\varepsilon_1=2$ and $\varepsilon_2=10$. In the first case ({\sc case I} which is a match composite), the conductivities are chosen as $\sigma_1=10^{-12}\ \siemens\per\meter$ and $\sigma_2=10^{-10}\ \siemens\per\meter$. In the other case ({\sc case II} which is a reciprocal composite), the conductivities of the phases are interchanged, $\sigma_1=10^{-10}\ \siemens\per\meter$ and $\sigma_2=10^{-12}\ \siemens\per\meter$. 

The results for {\sc case I} and {\sc case II} are illustrated with open ($\circ$) and filled ($\bullet$) symbols in Figure~\ref{fig:hig}. In Figure~\ref{fig:hig}a, permittivity values obtained from triangular lattice arrangement $\epsilon_{\rm h}$ are plotted against the permittivities obtained for the square lattice arrangement $\epsilon_{\rm s}$. There is no linear cross-correlation between the calculated permittivities ($\circ$ symbols in the figure) for the two different topologies for match composites ({\sc case I}). This sort of behavior has also been reported elsewhere for simulations performed on $4\times4$ binary mixtures~\cite{Tuncer2002elec}. However, the scattered full symbols ($\bullet$) for reciprocal composites ({\sc case II}) indicate that there is a tendency of correlation since the distribution is like an ellipse. The ellipse is slightly parallel to $\epsilon_{\rm h}=\epsilon_{\rm s}$ diagonal relation but shifted down showing high frequency permittivity values for square lattice arrangement has higher polarization compared to the triangular arrangement. The regions of $>90\ \%$ of the observed values are also plotted in the figure as dashed circle and ellipse to display the differences between the cases.   

In Figure~\ref{fig:hig}b the normalized conductance values $\sigma$ at $2\pi\ \milli\hertz$ are shown. It is important to remind you that at this frequency value non-percolating regular structures exhibit clear ohmic losses. For the match composites ({\sc case I}), the distribution of $\sigma_{\rm h}$ versus $\sigma_{\rm s}$ is narrow compared to {\sc case II}. It is also remarkable that there is a clear boundary between match and reciprocal composites such that the boundary line is $\sigma_{\rm h}=\sigma_{\rm s}$. This characteristic coincides with that of~\citeasnoun{Tuncer2002elec} whose results illustrate that the topological differences can be obtained from electrical properties. 

\begin{figure}[t]
  \centering{
    {\includegraphics[height=\figtwo]{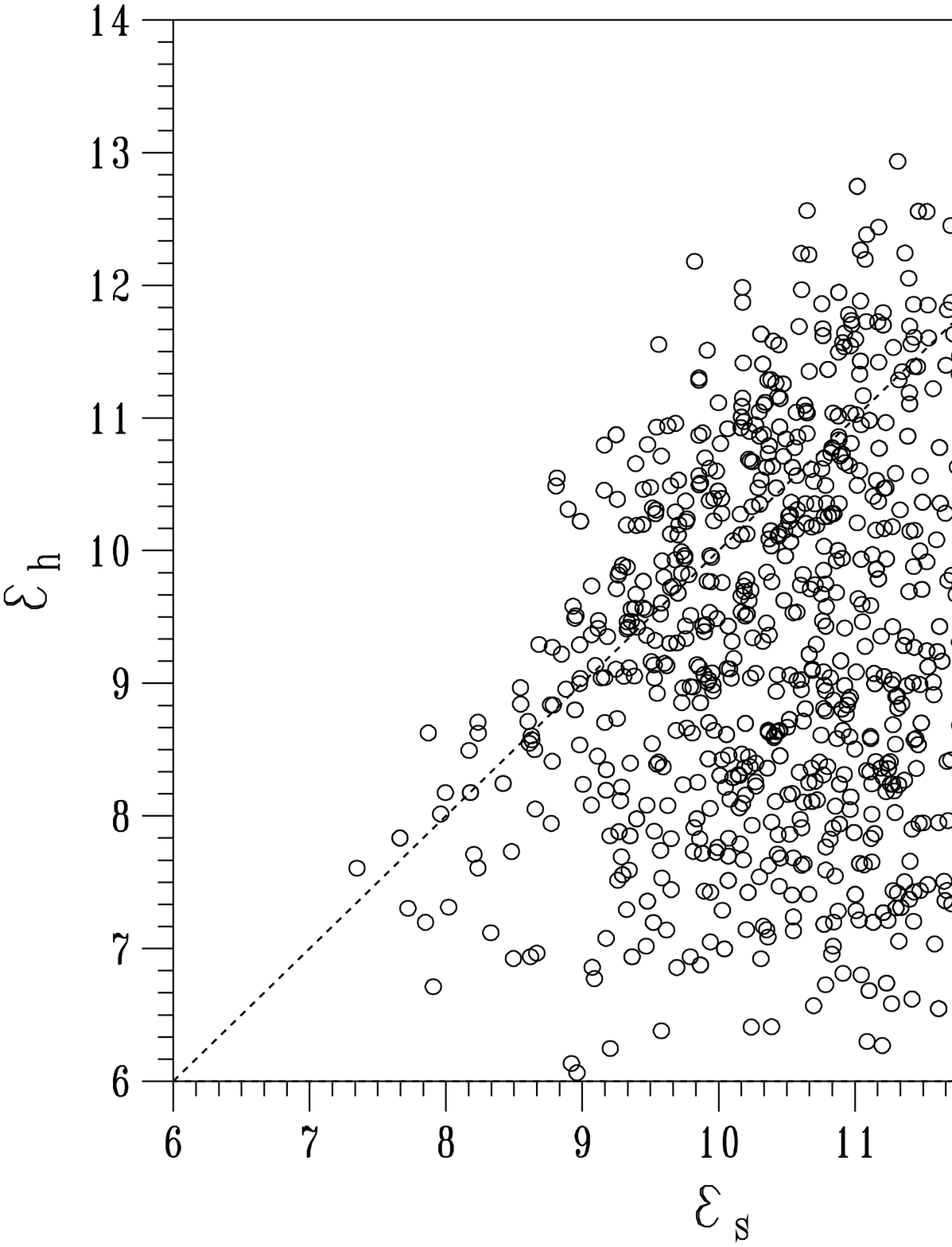}}
    {\includegraphics[height=\figtwo]{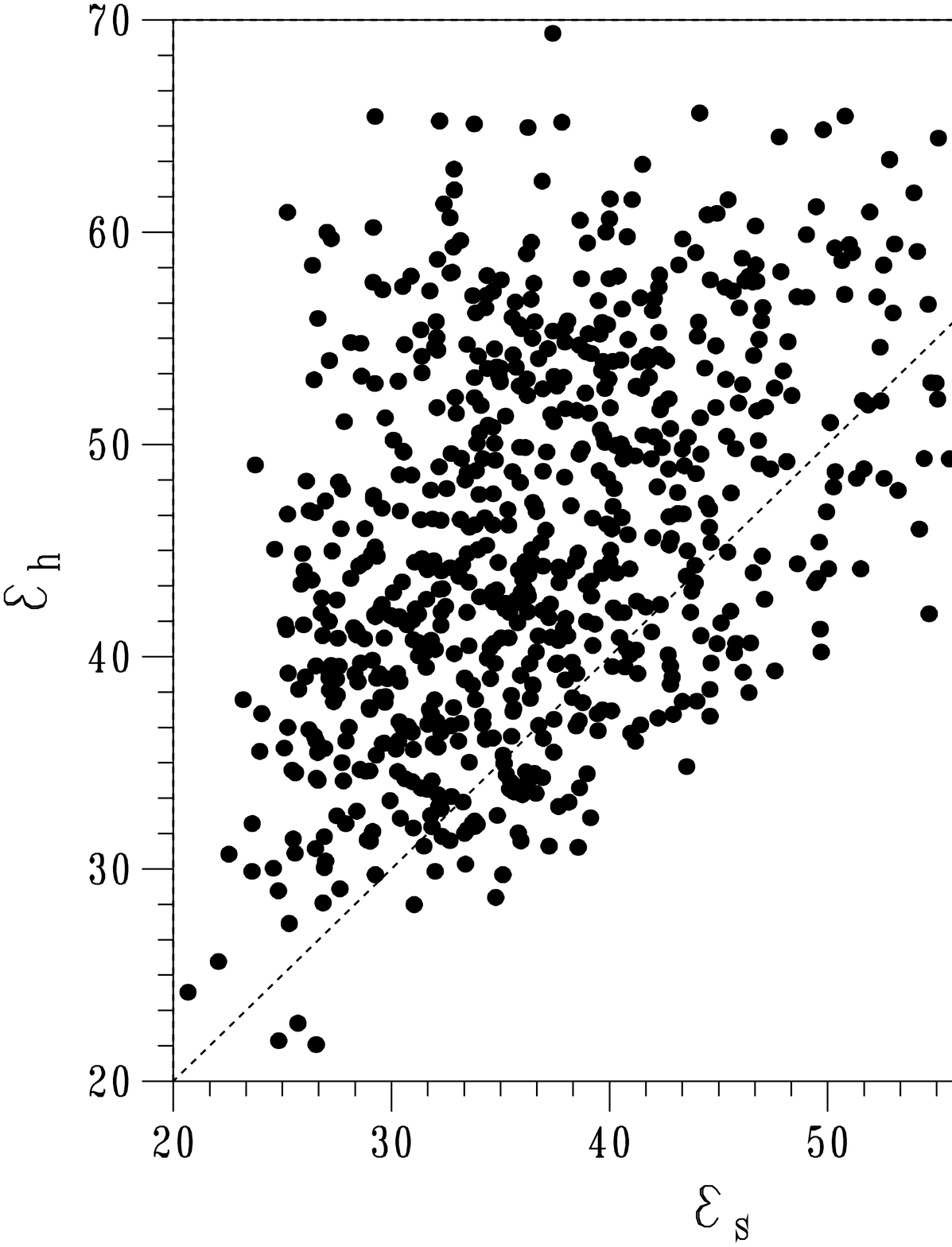}}
    }
  \caption{Relation between permittivities, $\varepsilon\equiv\epsilon+\chi(0)$ at $\omega=2\pi\ \milli\hertz$ for the disordered structures with different networks, (a) when $\epsilon_1<\epsilon_2$ and $\sigma_1<\sigma_2$ ($\circ$) and (b) when $\epsilon_1<\epsilon_2$ and $\sigma_1>\sigma_2$ ($\bullet$). Subscripts $h$ and $s$ represent the honeycomb and square tiles.  \label{fig:low}}
\end{figure}

At low frequencies the differences between the dielectric permittivities of the considered cases are more significant. In Figures~\ref{fig:low}a and \ref{fig:low}b, the dielectric permittivities at $2\pi\ \milli\hertz$ are presented for match and reciprocal composites, respectively. In Figure~\ref{fig:low}a the influence of structural differences leads higher low frequency permittivity values, $\varepsilon=\epsilon+\chi(0)$, for the square lattice arrangement than triangular lattice with honeycomb tiles. When the conductivities of the constituents are interchanged, the distribution of $\varepsilon_{\rm h}$ versus $\varepsilon_{\rm s}$ becomes upside-down compared to the {\sc case I} as presented in Figure~\ref{fig:low}b. The obtained dielectric strengths $\chi(0)$ are much higher for the reciprocal composites $15<\chi(0)<65$ than the match composites  $1<\chi(0)<9$. It is worth to mention that the biased distributions in Figure~\ref{fig:low} indicate lower and upper bounds from the structural differences between two lattices; an upper bound for match composites and a lower bound for reciprocal composites, respectively.

%The match composites ({\sc case I})  results in lower dielectric permittivities, $\varepsilon=\epsilon+\chi(0)$, than the reciprocal ones ({\sc case II}).  It is interesting to observe that the distributions are biased, and they swith character when the conductivities are altered. When electrical properties of phases are chosen such that the phase with the higher permittivity has also the higher conductivity of two phases, the topological differences resulted a lower dielectric permittivities for the honeycomb tiled triangular lattice. However, for the other considered case, {\sc case II}, the opposite is true. 

\subsection{Frequency dependent dielectric properties}

\begin{figure}[tp]
  \centering{
    {\includegraphics[angle=-90,width=\figtwob]{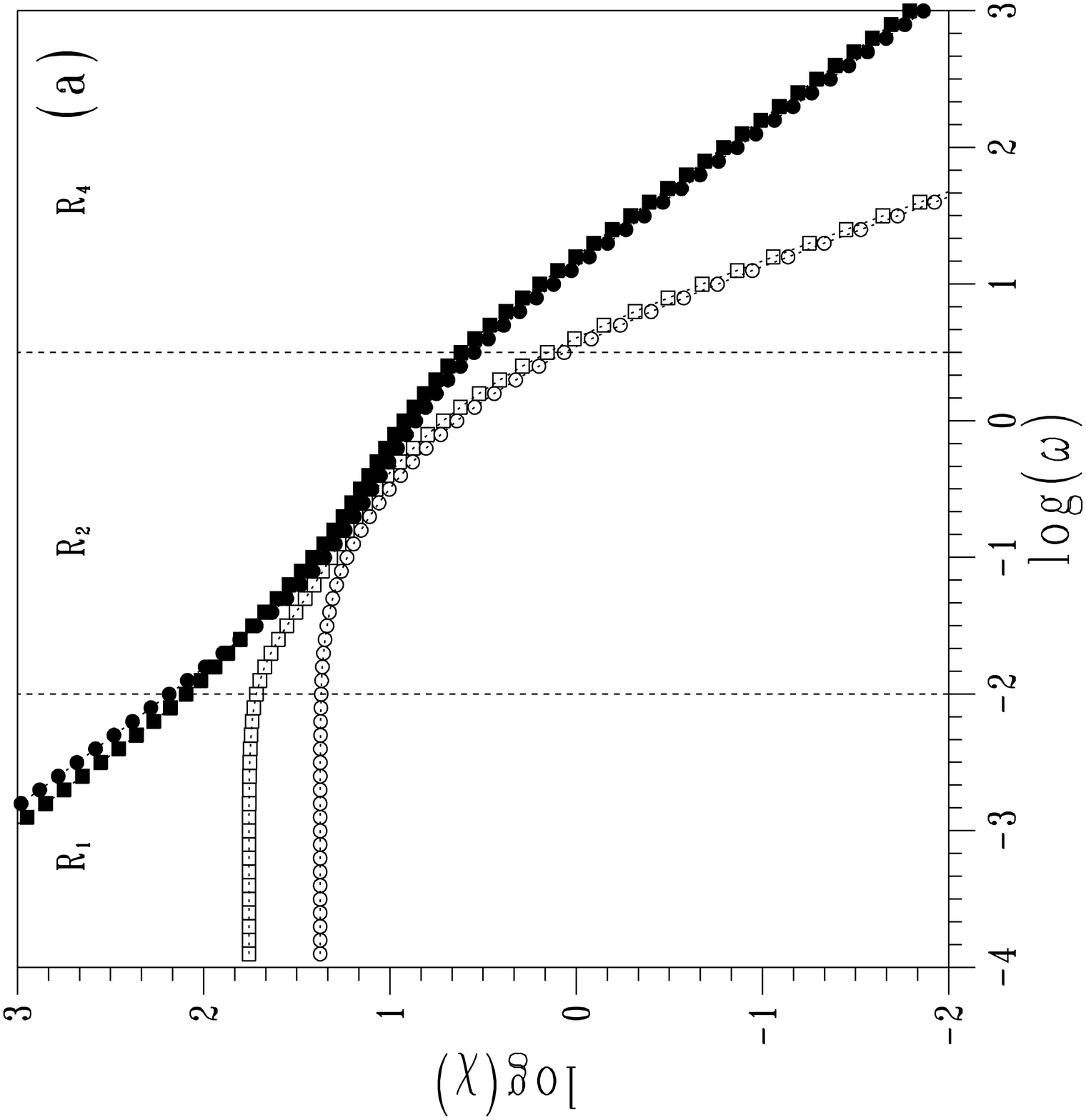}}
    {\includegraphics[angle=-90,width=\figtwob]{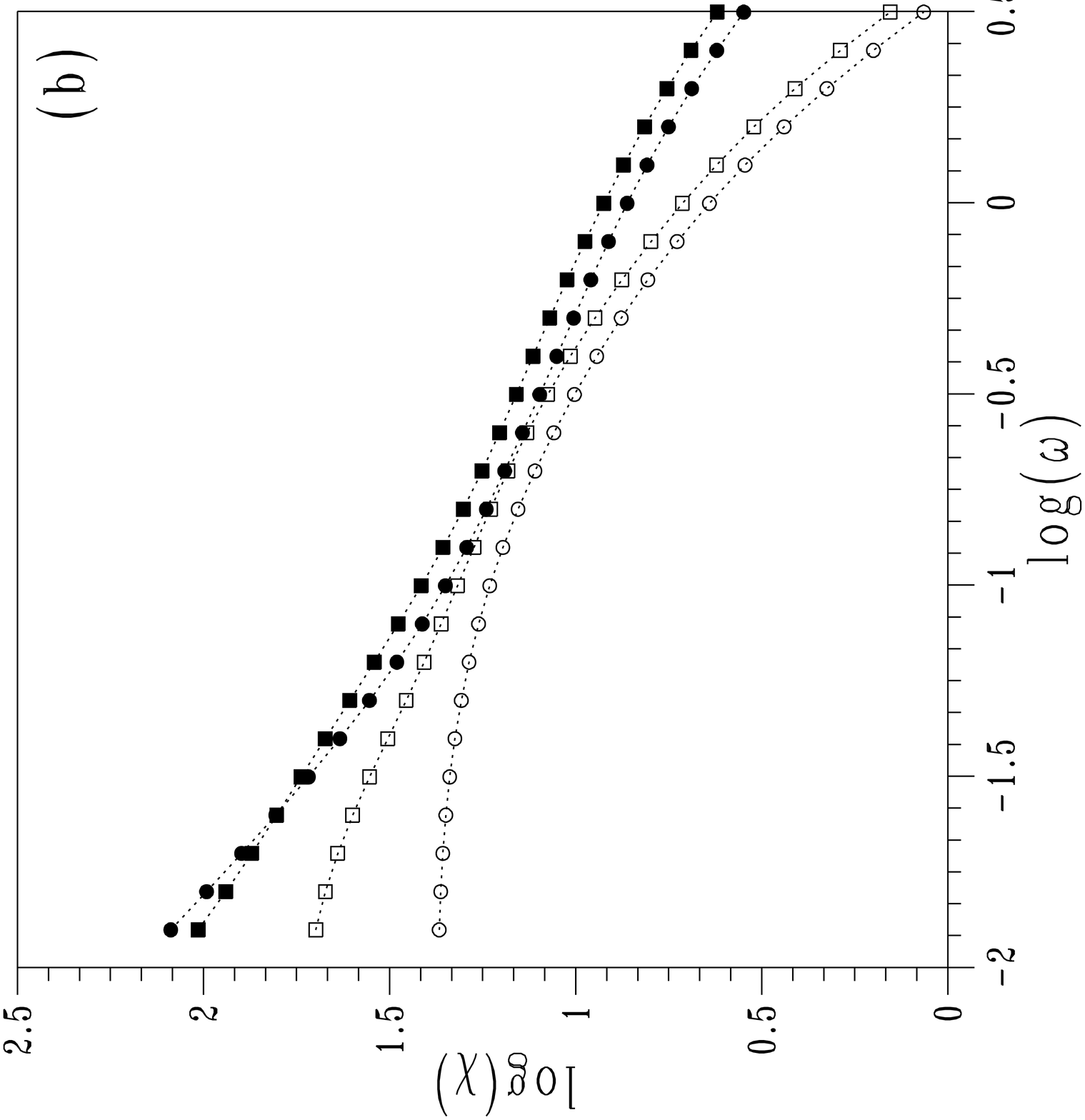}}\\
    {\includegraphics[angle=-90,width=\figtwob]{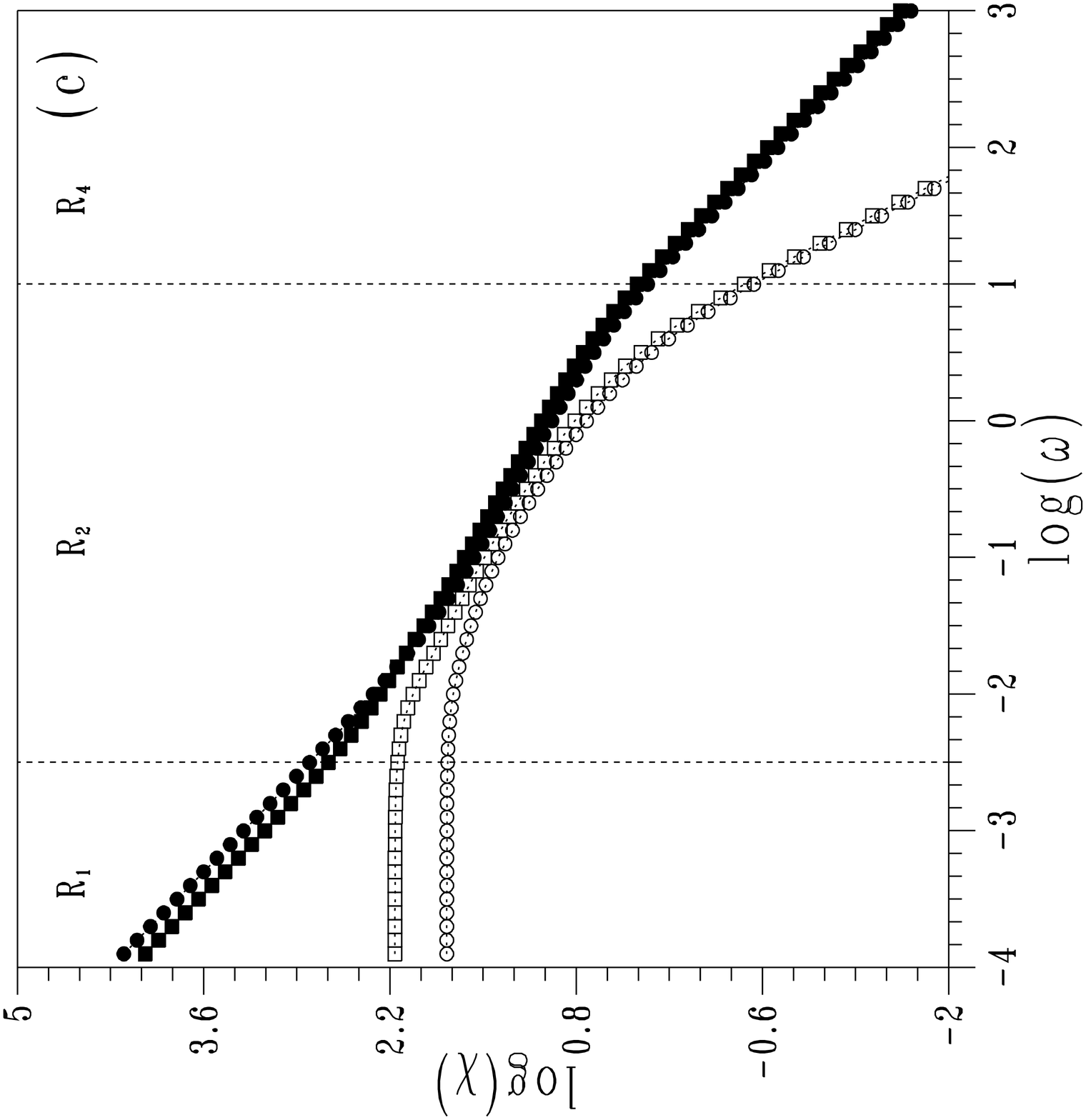}}
    {\includegraphics[angle=-90,width=\figtwob]{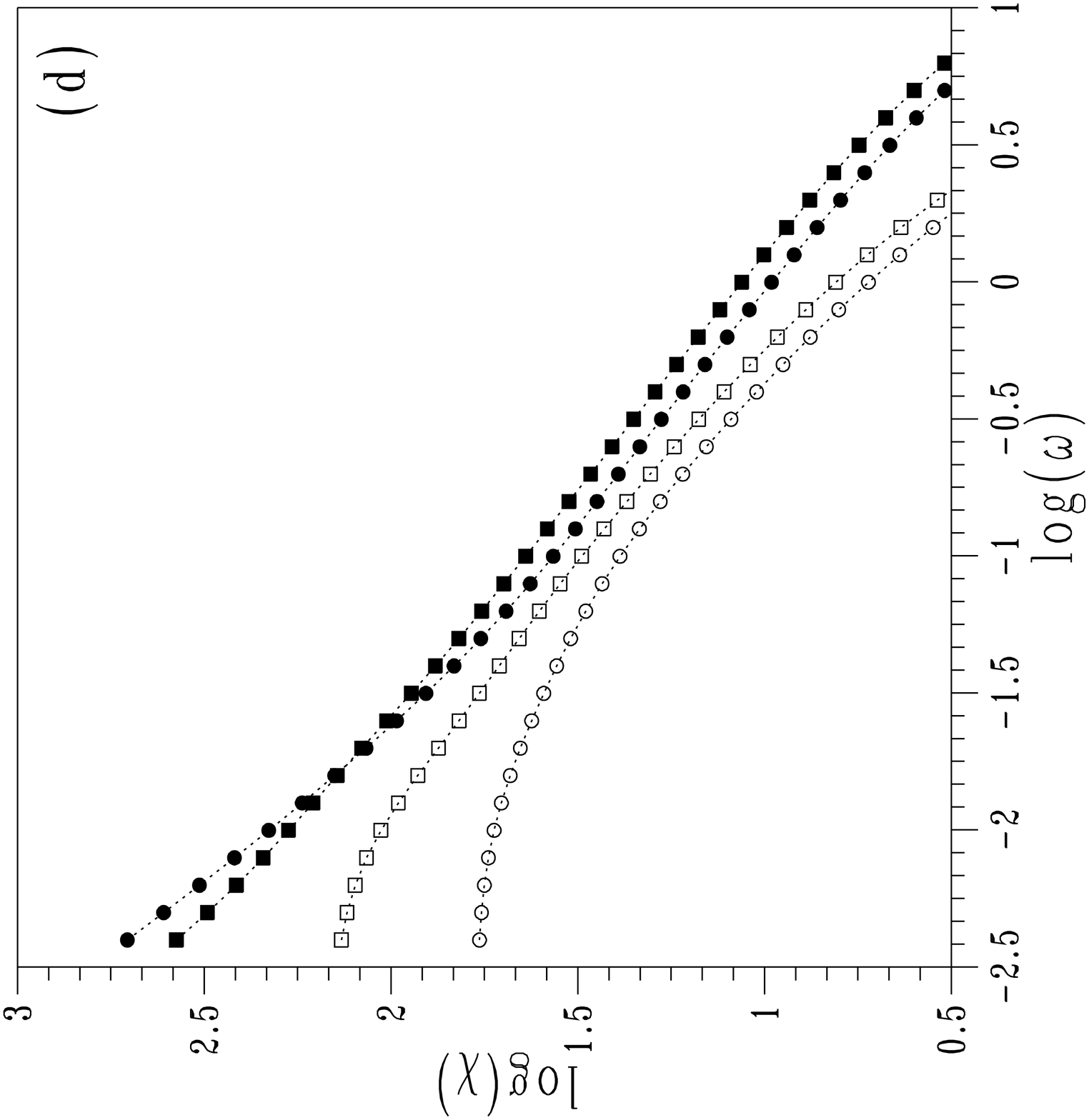}}
    }
  \caption{(a) and (b) Dielectric responses obtained for two chosen random structures with square ($\circ-\bullet$) and honeycomb ($\square-\blacksquare$) tiles. The regions presented as $R_2$ are enlarged in (b) and (d) to emphasize the presence of the low frequency dispersions. The open ($\circ-\square$) and filled ($\bullet-\blacksquare$) symbols represent the real, $\chi'$ and imaginary, $\chi''$ parts of the calculated dielectric susceptibility.\label{fig:res6}}
\end{figure}

The frequency dependent complex dielectric susceptibilities of two selected structures with different lattices are presented in Figure~\ref{fig:res6}. The electrical properties of the phases are chosen such that $\epsilon_1<\epsilon_2$ and $\sigma_1>\sigma_2$ similar to those reciprocal composites in the previous section. We mark regions in which the relaxation changes its behavior as in Figure~\ref{fig:lfd}. The Figures \ref{fig:res6}b and \ref{fig:res6}d are enlargements of the region $R_2$ in sub-figures (a) and (c). The calculated dielectric data do not show the presence of a region ($R_3$ in Figure~\ref{fig:lfd}) with low frequency dispersion satisfying the condition in Eq. (\ref{eq:cond2}). However, we include another region at high frequencies, $R_4$ which shows initiation of relaxation processes. At this region both structures have similar dielectric relaxations with $\chi'(\omega)\propto\omega^{-2}$ and $\chi''(\omega)\propto\omega^{-1}$ indicating the relaxations are Debye-type as shown in Figure~\ref{fig:res6}a and \ref{fig:res6}c. However, the dispersions in the intermediate region, $R_2$, which is identical to the region presented in Figure~\ref{fig:lfd} can be expressed as distribution of relaxation times~\cite{Tuncer2000b,Tuncer2002b}. There is only slight differences between two lattice structures at high frequency end of the data in $R_2$, but at the low frequency side both the structural effects are distinguishable from the real $\chi'$ and the imaginary $\chi''$ parts of the  susceptibilities. The distributions presented in Figure~\ref{fig:hig}b and \ref{fig:low} are once more significant such that at the lowest possible frequencies (the region $R_1$), the structures with honeycomb tiles have higher $\chi'$, on the other hand, lower $\chi''$ than square tiles. 
The structural changes influence the range of the low frequency dispersion observed in $R_2$. The dc conduction contributions are significant only in $R_1$ region. 

The dielectric response in Figure~\ref{fig:res6} has a power law behavior, $\chi(\omega)\propto(\imath\omega)^{-a}$ with $a=[0.5,0.55]$ in a short frequency region, $0.1\ \hertz<\omega<0.3\ \hertz$. When $a=0.5$ the electrical process is associated with the semi-finite or finite diffusion process~\cite{MacDonald1987}. Therefore, the {\sc mws} polarization lead a low frequency dispersion. %The dielectric behaviors in $R_2$ are enlarged and displayed in Figures~\ref{fig:res6}b and \ref{fig:res6}d to show the differences in the amplitutes of the real and imaginary parts of the suscepribilities, $\chi(\omega)$. The structural differences are clear and the structures with honeycomb tiles have higher $\chi(0)$ values than the structures with square tiles as presented in Fig~\ref{fig:low}b. 
{\em As a result a condition for the `normal' low frequency dispersion is created here with a slightly reciprocal composite}. 

\begin{figure}[tp]
  \centering{
    {\includegraphics[angle=-90,width=\figtwob]{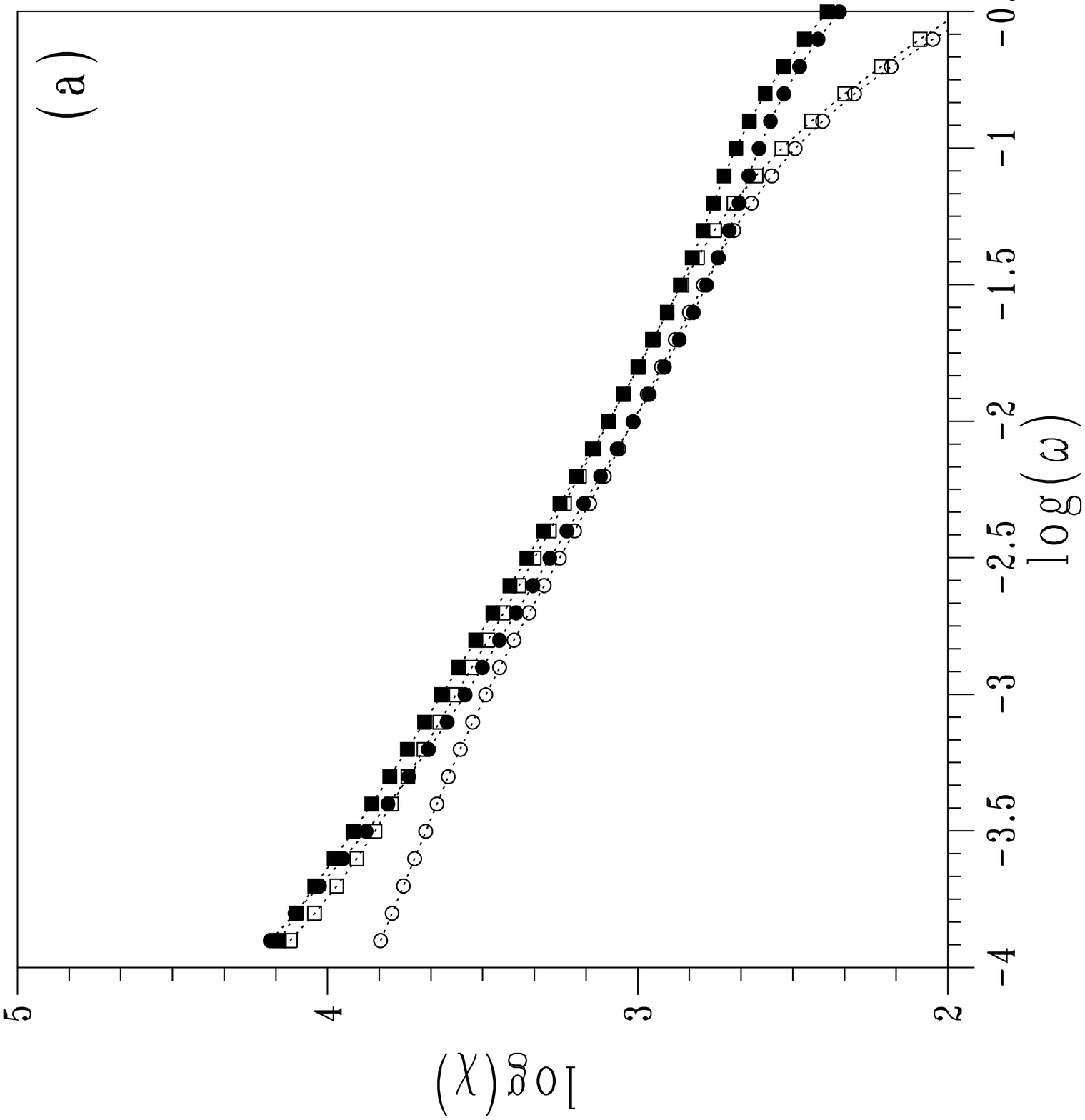}}
    {\includegraphics[angle=-90,width=\figtwob]{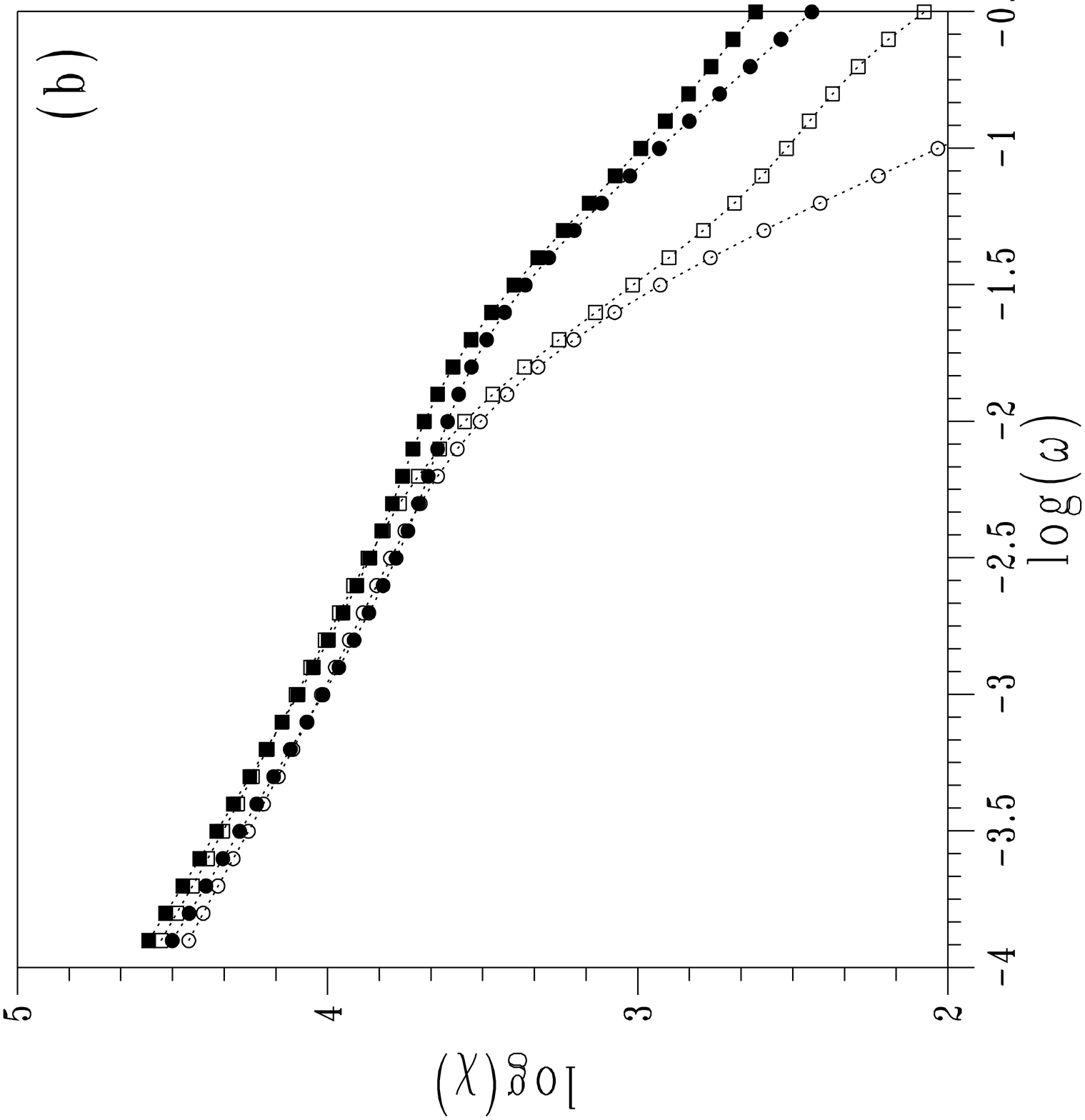}}\\
    {\includegraphics[angle=-90,width=\figtwob]{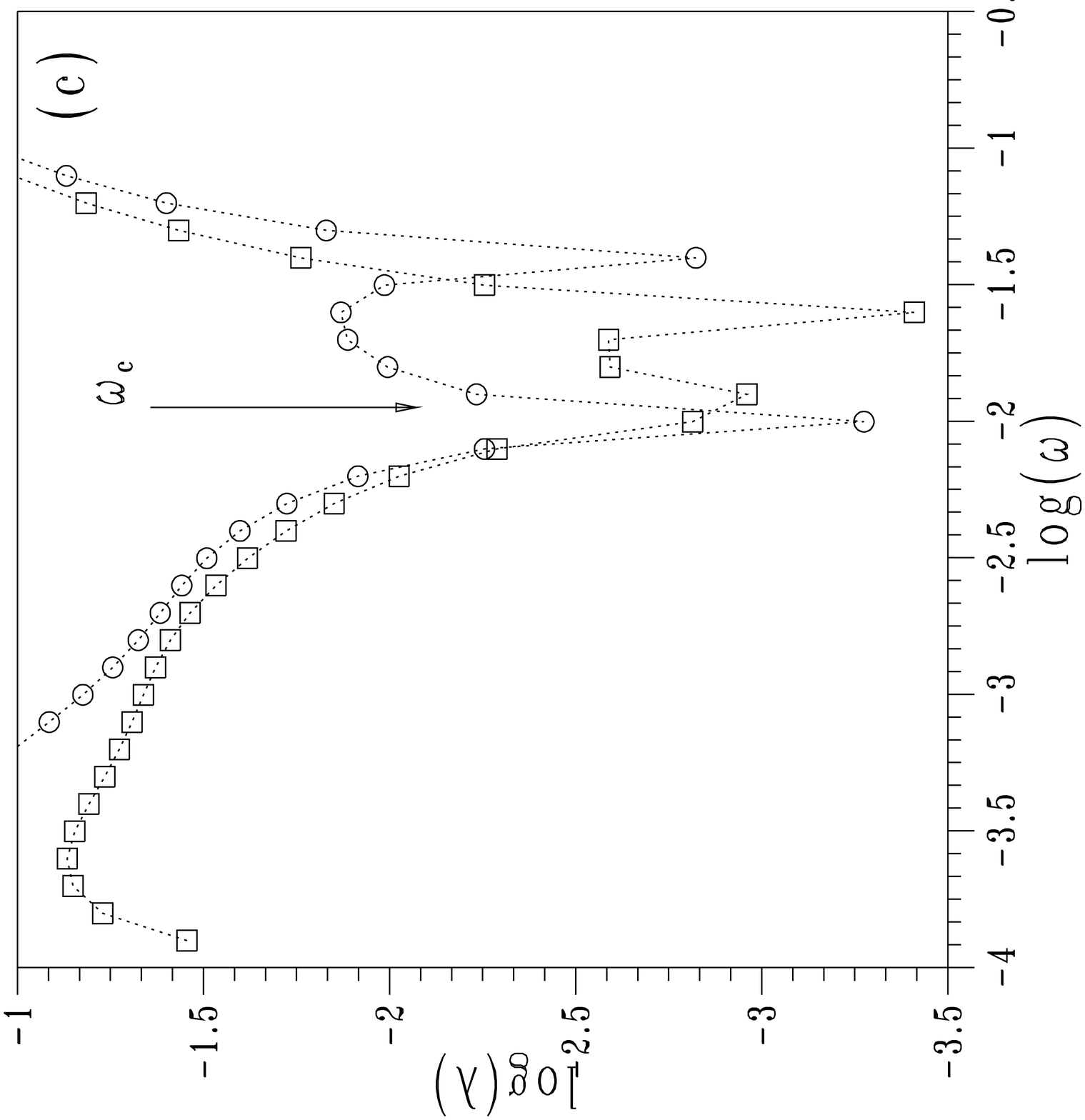}}
    {\includegraphics[angle=-90,width=\figtwob]{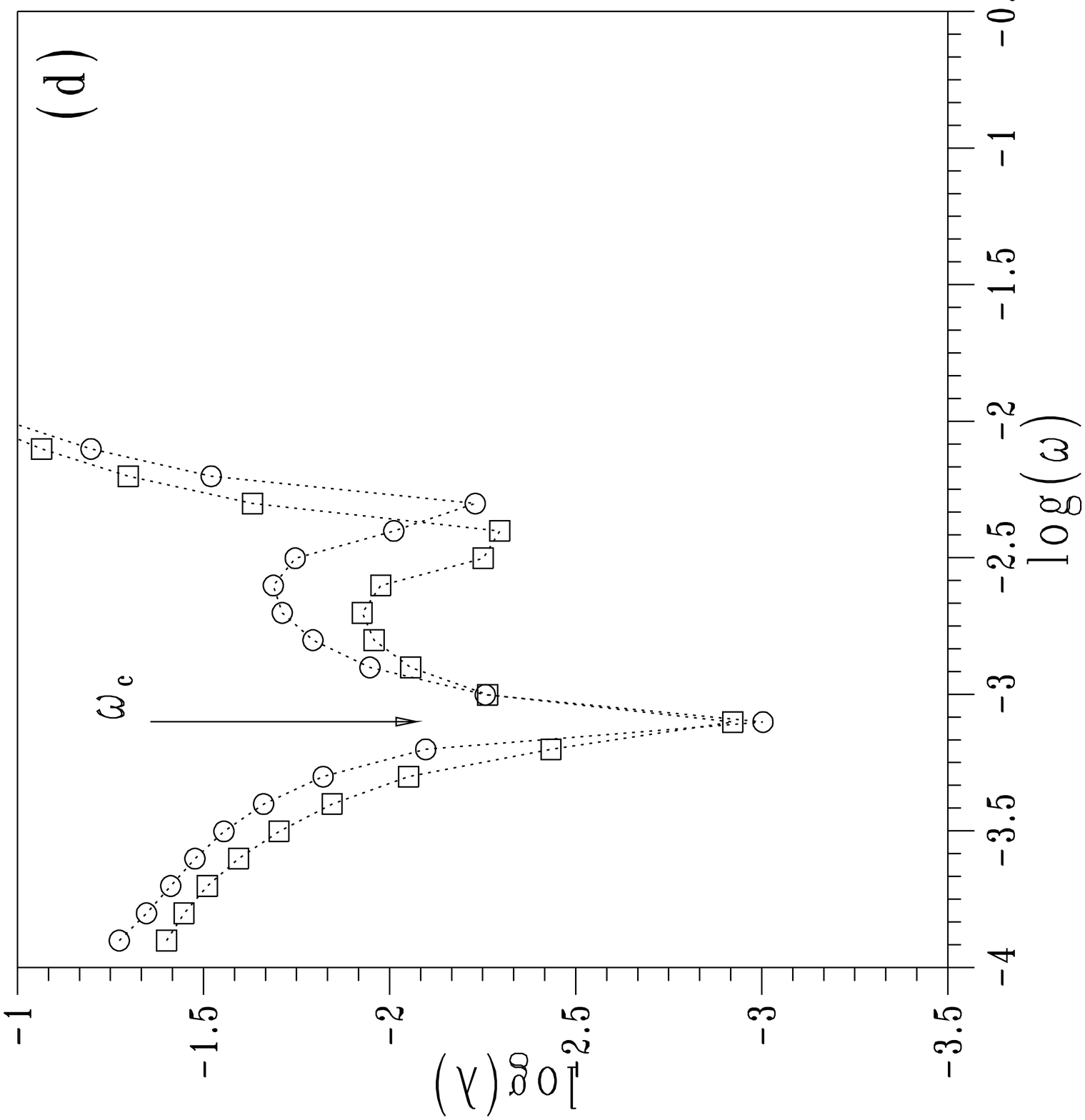}}
    }
  \caption{Dielectric responses of random structures by keeping phase parameters of phase (1) constant, $\epsilon_1=2$ and $\sigma_1=10^{-10}\ \siemens\per\meter$, and assuming phase (2) parameters as (a) $\epsilon_2=2000$ and $\sigma_2=10^{-14}\ \siemens\per\meter$, and (b) $\epsilon_2=20000$ and $\sigma_2=10^{-15}\ \siemens\per\meter$. The open ($\circ-\square$) and filled symbols ($\bullet-\blacksquare$) represent the real, $\chi'(\omega)$, and imaginary, $\chi''(\omega)$, parts of the dielectric susceptibilities. (c) and (d) are plots of the ratio of real, $\chi'(\omega)$, and imaginary, $\chi''(\omega)$, parts of the dielectric susceptibilities for the considerations in sub-figures (a) and (b), respectively. The open circle ($\circ$) and open box ($\square$) in (c) and (d) represent the $\lambda$ values obtained for square and honeycomb tiles, respectively.  \label{fig:res7-8}}
\end{figure}

To have the ``anomalous'' low frequency dispersion,  the parameters of phase (1) are kept constant, $\epsilon_1=2$ and $\sigma_1=10^{-10}\ \siemens\per\meter$, and the parameters of phase (2) are, respectively, chosen (i) $\epsilon_2=2000$ and $\sigma_2=10^{-14}\ \siemens\per\meter$, and (ii) $\epsilon_2=20000$ and $\sigma_2=10^{-15}\ \siemens\per\meter$ which is an extreme reciprocal composite. The simulated dielectric susceptibilities are presented in Figures~\ref{fig:res7-8}a and \ref{fig:res7-8}b. In contrast to the previous assumption in Figure~\ref{fig:res6}, the $R_2$ region in this extreme reciprocal composite increases in size even more as displayed in Figure~\ref{fig:res7-8}. This leads to unobservable ohmic losses which are present at lower frequencies. Moreover, the flattening of the real part (the finalization of polarizations) in $R_1$ is not observable in the considered frequency range in the similar region as in Figure~\ref{fig:res6}a and \ref{fig:res6}c. As expected, the $\chi(0)$ and $\epsilon$ values are higher than the previous calculations. At the same time, again the power law behavior in the region $R_2$ is in $\chi(\omega)\propto(\imath\omega)^{-a}$ form however with a slightly higher exponent, $a=[0.5,0.6]$, and satisfying the condition in Equation (\ref{eq:cond2}). Yet, no sign of region $R_3$ is noticeable to the naked eye which would indicate the anomalous low frequency dispersion. 
%%%%%There are significant low frequency dispersions in the figures. Moreover, the range of frequency, in which the low frequency dispersions are observed, are different. 
To show the presence of the anomalous behavior, a parameter, $\lambda$, is assigned,
\begin{equation}
  \label{eq:lambda}
  \lambda=\left|\log\left[\frac{\chi'(\omega)}{\chi''(\omega)}\right]\right|
\end{equation}
This parameter converges to zero as we approach the critical frequency $\omega_c$ in Figure~\ref{fig:lfd}. The logarithm of $\lambda$ versus frequency $\omega$ is shown in Figures~\ref{fig:res7-8}c and \ref{fig:res7-8}d for the dielectric responses at Figures \ref{fig:res7-8}a and \ref{fig:res7-8}d, respectively. The first local minimum at high frequency is the first cross-over of $\chi'$ and $\chi''$ after the start of the polarization. The region between the local minimums in Figures~\ref{fig:res7-8}c and \ref{fig:res7-8}d indicate frequency interval $R_3$ in Figure~\ref{fig:lfd}, in which the condition (\ref{eq:cond1}) is satisfied. The critical frequencies, $\omega_c$, for the anomalous low frequency dispersions are marked with arrows in the figures. Although, the change of the exponent of the power law is not large before and after $\omega_c$, as stated in the literature~\cite{HillLFD}, including (i) extra phases in considered structures or (ii) using phases with low frequency dispersions or (iii) using a higher computation size than $16\times16$ might alter these exponents. 

The influence of topology on the  observed low frequency dispersion behavior in the middle part of the responses is not that significant, however, on both sides of the frequency window, the responses are influenced by the considered lattice structures.
Finally, since the assumed systems in our simulations are small in size ($16\times16$) and have the finite smallest tile sizes, a dielectric relaxation of Debye-like behavior with a single-time constant is present at high frequencies, the $R_4$ region. This characteristics has also discussed by~\citeasnoun{MacDonald1987} as the high- and low-frequency extreme limits of responses. It has been also presented in~\citeasnoun{Tuncer2002b}, in which the wide distribution of relaxation times should start with a single-time constant dielectric relaxation~\cite{Tuncer2002b}. Above both frequency extremes the low frequency dispersion behavior fails. The experimental evidence of this kind of responses in real materials, similar to the dielectric relaxations presented in Figure~\ref{fig:res7-8}a and \ref{fig:res7-8}b have been reported in the literature~\cite{Jonscher1983,HillLFD}.
\begin{figure}[t]
  \centering{
    {\includegraphics[height=\figtwo]{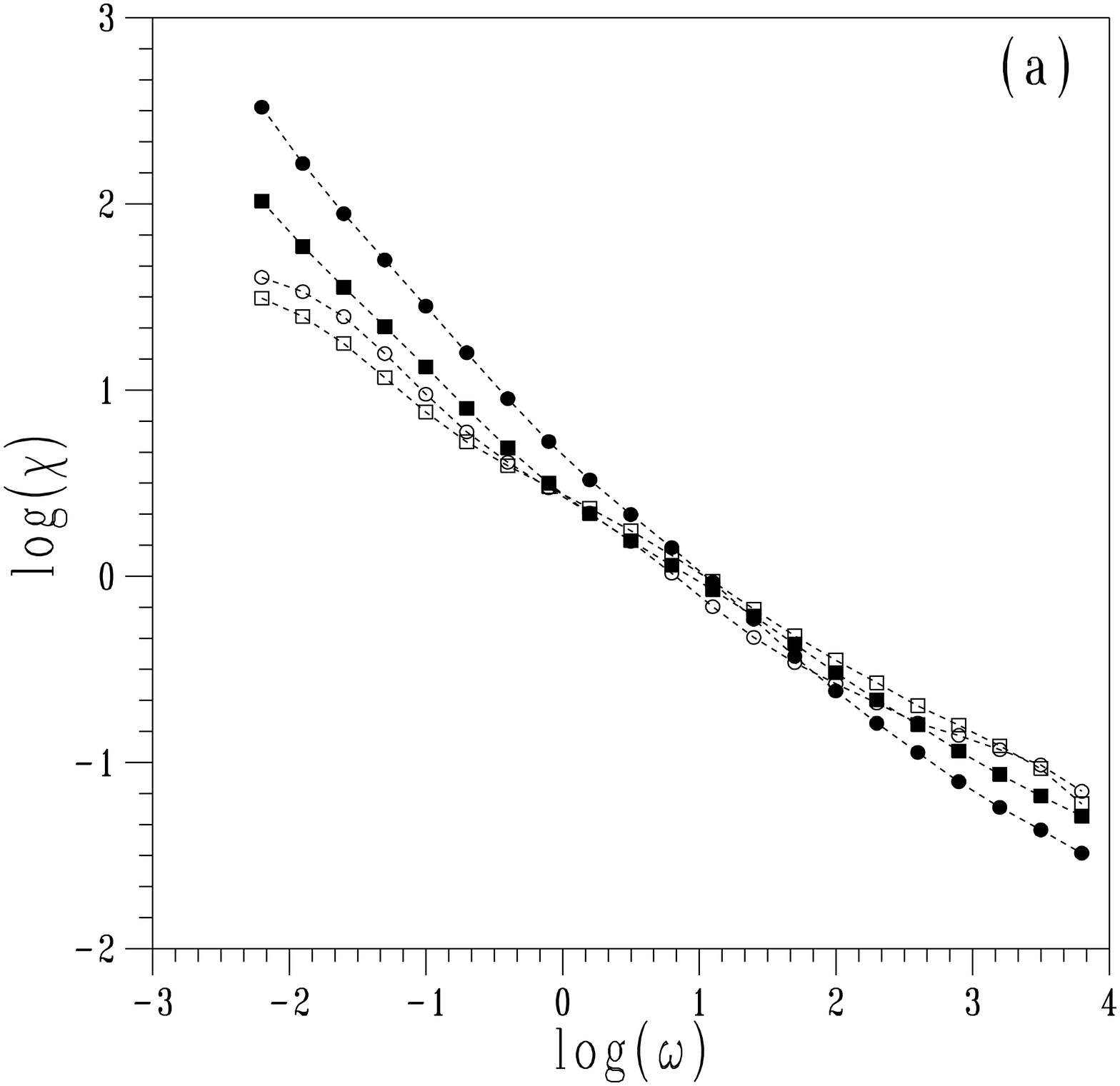}}
    {\includegraphics[height=\figtwo]{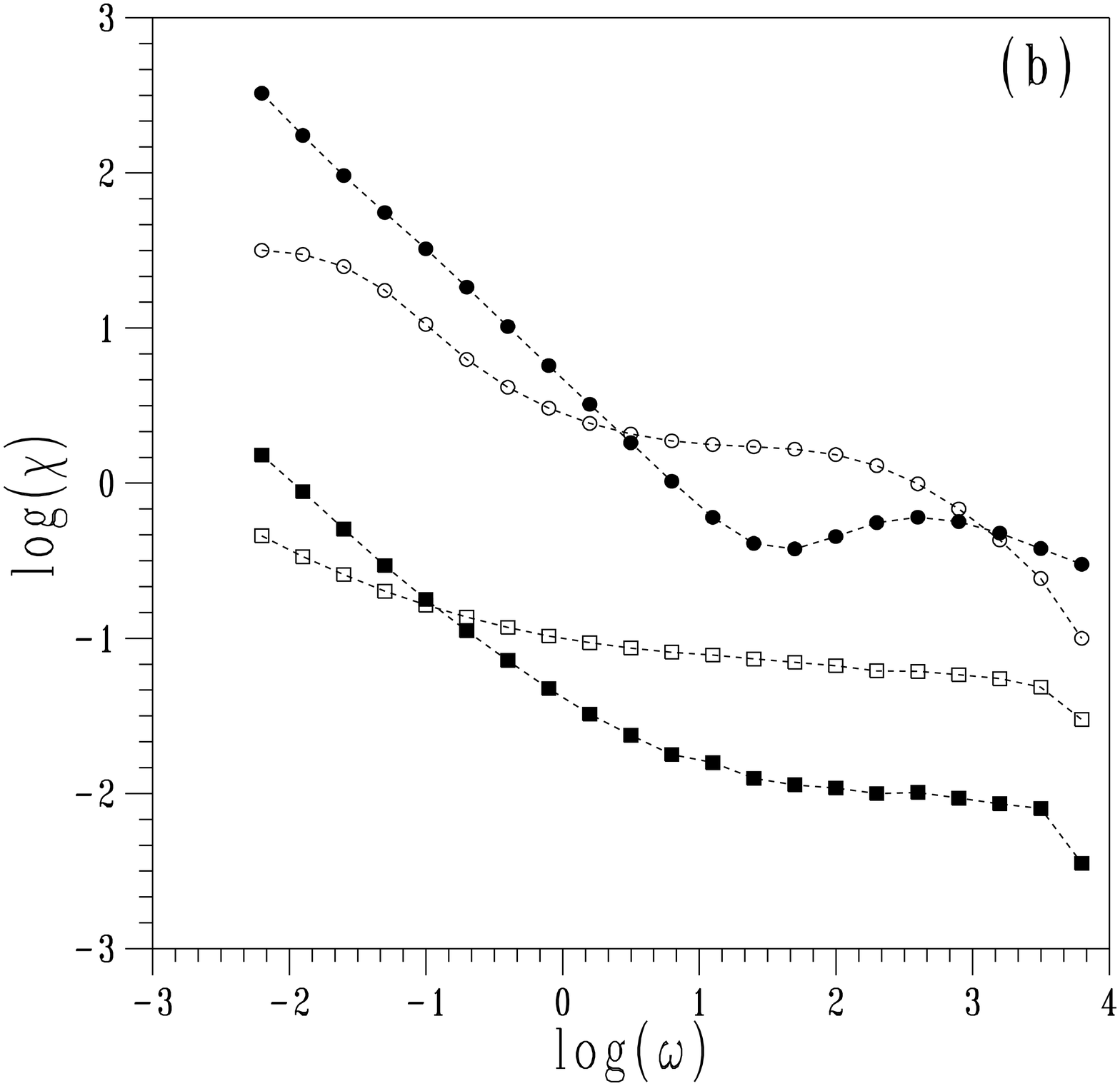}}
    }
  \caption{Dielectric responses of silicone rubber composites (a) with Aluminum Trihydrate powder (b) and glass-beads as inclusions. The open ($\circ-\square$) and filled symbols ($\bullet-\blacksquare$) represent the real, $\chi'(\omega)$, and imaginary, $\chi''(\omega)$, parts of the dielectric susceptibilities. In sub-figure (a) $\chi(\omega)$ of materials with the same formulation are presented. In sub-figure (b), the square ($\square-\blacksquare$) and circle ($\circ-\bullet$) symbols represent the the  materials with small and large glass-beads, respectively.  \label{fig:mes1}}
\end{figure}

\subsection{Dielectric response of filled systems}
In Figure~\ref{fig:mes1}, dielectric susceptibilities, $\chi(\omega)$, of silicone polymer based composite materials are presented. In the sub-figures, (a) samples prepared with 50 weight \% Aluminum Trihydrate powder which has arbitrary-shapes, and (b) with glass-beads which are spheroidal are displayed. When the samples with Aluminum Trihydrate powder are considered the dielectric responses, presented in Figure~\ref{fig:mes1}a, are similar to those presented in Figure~\ref{fig:res6} of the previous section. In this sample, therefore, the conductivity of the filler material (phase 2) might have been considered to have a lower conductivity value than the matrix phase (phase 1), $\sigma_2<\sigma_1$, and the permittivity of the filler  has a higher value than the silicone rubber, $\epsilon_2>\epsilon_1$, a reciprocal composite. The data from~\citeasnoun{Tuncer2000a} show that both these conditions are true, however, the ratio of permittivities $\epsilon_1/\epsilon_2$ is not as high as we consider in this paper. The ratio of the conductivities of the phases is, on the other hand, not an easy task to assign. However, the data obtained from different composites with varied concentrations of filler indicate that for some concentration value (less than 50 weight \%) conductivity of the composite decreases. This proves that the conductivity of filler $\sigma_2$  is lower than the matrix when effective medium theories are applied~\cite{SihvolaBook}. Therefore, the observed low frequency dispersions in Aluminum Trihydrate filled silicone rubber samples could be from the {\sc mws} polarization. 

In  Figure~\ref{fig:mes1}b, two responses, in which the composites contain two different sizes of glass-beads are illustrated. The larger glass-beads have higher conductivity compared with the small glass-beads. We presume that the conductivity of the polymer is lower than the conductivity of the large glass-bead, and higher than the small glass-beads from the dielectric spectroscopy measurements performed on the polymer. The material with small glass-beads shows an anomalous low frequency dispersion (a reciprocal composite), however, the material with large glass-beads indicates the presence of two dielectric relaxations and a dc conductivity in the complex susceptibility (a match composite). These experimental results together with the numerical simulations demonstrate that low frequency dispersion behavior observed in dielectric mixtures can be also modeled using computer simulations on disordered structures just by assuming the {\sc mws} polarization--none intrinsic polarization of constituents.

More illustrative examples in which higher dielectric constants have observed are porous materials that are saturated with conductive phases such as salts and water~\cite{Bo1996,Papathan2000,Papathan2001,Capaccioli3,shen1990}. In these systems real part of the permittivity are reported to reach even $10^7$ levels as in~\citeasnoun{Bo1996}, however in those system the liquid conductivity was responsible for the larger polarization observed in the system. In the present work the generated topologies were similar to cross-section of porous materials and by assigning material parameters which satisfied the condition for ``reciprocal mixture'', large dielectric permittivity values can be obtained as in liquid-solid mixtures.

\section{Conclusions}

The performed {\sc fem} calculations on binary disordered dielectric mixtures indicate that the topology of the system is as important as the intrinsic material properties of the constituents to characterize the system. In the simulations $16\times16$ tiles are generated on square and triangular lattices with the same tile arrangement. 
 
When a match composite, $\epsilon_1<\epsilon_2$ and $\sigma_1<\sigma_2$,  is taken into consideration, the topological differences point out no significant relation between the considered topological arrangements for permittivity values at high frequencies. This is due to the low value of the ratio of the intrinsic permittivity values of the phases. For reciprocal composites, $\epsilon_1<\epsilon_2$ and $\sigma_1>\sigma_2$, the structural differences does not influence the effective dielectric permittivities, $\epsilon$, of the mixture at higher frequencies than the characteristic interfacial polarization is observed. The relation between $\epsilon$-values are nearly-linear.  However, other calculated parameters, permittivity at low frequencies and dc conductivity, show influence of the structural differences for both match and reciprocal composites. This arises from the considered conductivity values which yield a high ratio of conductivities. The conductivity also affects the interfacial polarization which is visible at very low frequencies. This result confirms that permittivity and conductivity can be used as a tool to get structural information confirming the results of~\citeasnoun{Tuncer2002elec}. At frequencies much lower than the characteristic interfacial polarization is observed, the topological differences and the conductivity of the phases produce significant differences in the permittivities, $\epsilon+\chi(0)$. Depending on whether the conductivity of the second phase, $\sigma_2$, is higher or lower than $\sigma_1$, the dielectric strength of the interfacial polarization, $\chi(0)$, differs for the two structures, in which in one case, $\sigma_1>\sigma_2$, the structures with honeycomb tiles have higher polarizations, and the opposite is true for structures with square tiles when $\sigma_1<\sigma_2$. This indicated the importance of the connectivity in the mixture. 

Finally, by altering the conductivity and permittivity of the second phase while keeping the electrical properties of the first phase constant, we show that normal and anomalous low frequency dispersions can be obtained in dielectric mixtures with frequency independent permittivity and constant dc conductivities. The {\sc mws} polarization in these systems might be confused with those of normal and anomalous low frequency dispersions. The normal low frequency dispersion is observed for $\sigma_1\gg\sigma_2$ and $\epsilon_1<\epsilon_2$. In a like manner, the anomalous low frequency dispersion is obtained when $\sigma_1\gg\sigma_2$ and $\epsilon_1\ll\epsilon_2$. Dielectric responses of several polymeric composite systems obtained from experiments and of porous materials are compared with the responses calculated by the {\sc fem}. The observed low frequency dispersions in those systems might have been due to the {\sc mws} polarization and micro-structural disorder in the materials which create long paths for charge carriers which are observed as to polarization in the dielectric response.

\section*{References}
\bibliography{tuncer_low} 
\bibliographystyle{dcu}
%\end{pagewiselinenumbers}
\end{document}